\documentclass[journal]{IEEEtran}

\usepackage{graphicx, dblfloatfix}
\usepackage{amsmath}
\usepackage{amssymb}
\usepackage{graphics}
\usepackage{cite}
\usepackage{amsfonts}
\usepackage{textcomp}
\usepackage{multicol}
\usepackage{multirow}
\usepackage{color}
\usepackage{fixltx2e}
\usepackage{epstopdf}

\begin{document}

\title{Adaptive Polarization Control for Coherent Optical Links with Polarization Multiplexed Carrier }

\author{ Mehul Anghan, Nandakumar Nambath$^*$, Rashmi Kamran, and Shalabh Gupta\\
	Department of Electrical Engineering, IIT Bombay, Mumbai-400076 India\\
	$^*$School of Electrical Sciences, IIT Goa, Ponda-403401 India
	 \thanks{This paper has to be submitted for publication to a journal. A significant part of this work comes from \cite{Mehul} }}
\maketitle
\begin{abstract}
  Self-homodyne systems with polarization multiplexed carrier offer an LO-less coherent receiver with simplified signal processing requirement that can be a good candidate for high-speed short-reach data center interconnects. The practical implementation of these systems is limited by the requirement of polarization control at the receiver end for separating the carrier and the modulated signal. In this paper, effect of polarization impairments in polarization diversity based systems is studied and modeled. A novel and practical adaptive polarization control technique based on optical power feedback from one polarization is proposed for polarization multiplexed carrier based systems and verified through simulation results. The application of the proposed concept is experimentally demonstrated also for a QPSK system with polarization multiplexed carrier. 
\end{abstract}


\IEEEpeerreviewmaketitle

\section{Introduction}
\IEEEPARstart{M}{erit} of coherent modulation and demodulation techniques make these techniques suitable for communication through optical fibres at high data rates\,\cite{Coherent}. Dual polarization – Quadrature Phase Shift Keying (DP-QPSK) system has been commonly used for high data rates which utilizes diversity in both phase and polarization\,\cite{dpqpsk, ADP2}. This coherent technique uses a separate local oscillator (LO) at receiver and also requires a carrier phase recovery (CPR) and compensation module to overcome the effects of laser line-width and frequency offset between the transmitter and receiver lasers. Use of LO and CPR can be avoided in a polarization diversity based self-homodyne (SH) system, in which the carrier is polarization multiplexed with the modulated signal itself. Polarization impairments tied with optical system and channel causes mixing of the data in two orthogonal polarizations results in improper reception of message symbols. Polarization demultiplexing techniques like constant modulus algorithm and decision directed algorithm in electrical domain can be used to compensate these polarization impairments but it requires high speed signal processing \cite{EPOL, SPSH1}. Circuit implementation of high speed signal processing will be complex and power hungry. Manual or automatic polarization controller device (for example EPC-15-1-1-2 from Phoenix, EPC1000 from Novoptel) can be used to correct this kind of impairments. Some of the effects can be minimized by properly controlling the state of received polarization in optical domain itself\,\cite{PC1, pol}. Few polarization diversity based SH systems are demonstrated in prior work\,\cite{RPPMC11, 16QAM1}, although manual polarization controller has been used which is not a solution in practical scenario. In other work\,\cite{RPPMC2}, a polarization multiplexed carrier based SH system is demonstrated with direct detection that faces issue of IQ imbalance.

\begin{figure*}[!h]
	\centering
	\includegraphics[scale=0.34]{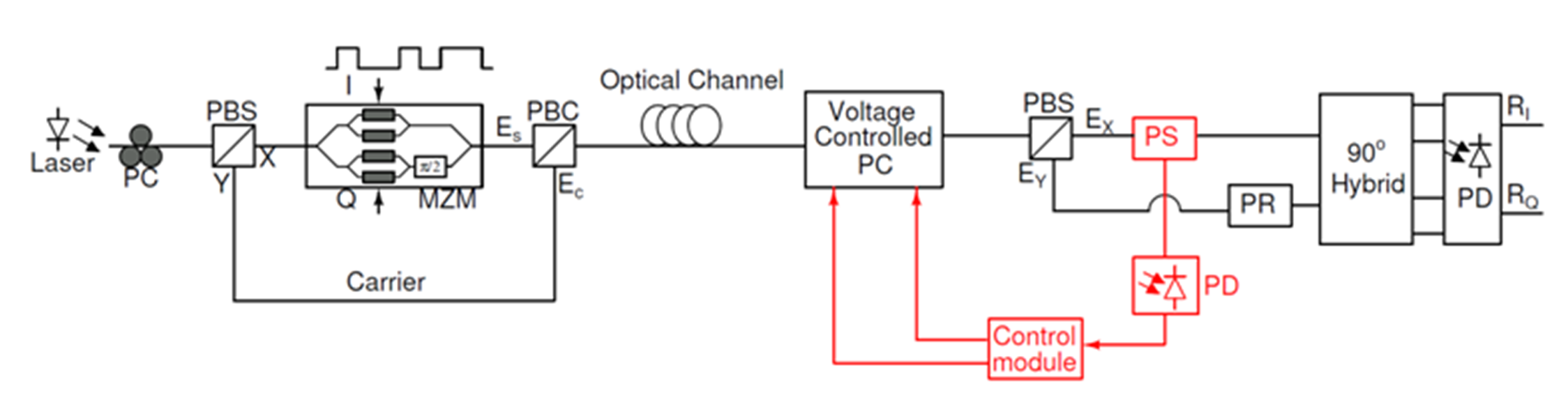}
	\caption{A polarization diversity based self homodyne system with proposed adaptive polarization control technique. PC: polarization controller, PBS/PBC: polarization beam splitter/combiner, PD: photo detector, PS: power splitter, and PR: polarization rotator.}
	\label{block}
\end{figure*}

Adaptive polarization control is found to be very useful for self-homodyne coherent optical links with polarization multiplexed carrier. Proposed adaptive polarization control for an polarization diversity based SH system is presented in Fig.\,\ref{block}, in which power of one of the polarization (after converting in electrical domain) is fed back to electronically controlled polarization controller.  In this paper, we have carried out modeling of SH systems for short reach links that include the effects of polarization impairments. A technique based on minimization of the optical power received in one of the polarizations, to control the state-of-polarization in short reach self-homodyne links, is presented with the use of feedback polarization control. A discrete time gradient descent based algorithm to achieve this minimization is presented and validated using simulations. The usefulness of polarization control has been experimentally verified for SH-QPSK system, where the proposed technique will work.

\section{Modelling of the polarization impairments:} 
\begin{figure}[!h]
	\vspace{-0.6cm}
	\centering
	\includegraphics[scale=0.3]{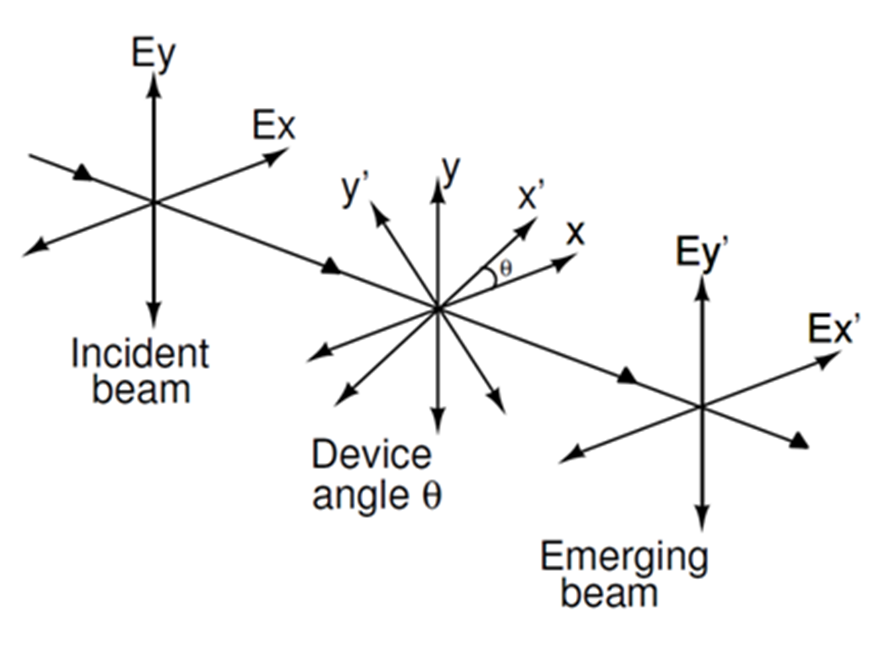}
	\caption{Effect on both polarizations with device angle.}
	\label{pol}
\end{figure}

Polarization impairments due to system components and fiber channel are discussed in this section. Polarization beam combiner (PBC) and polarization beam splitter (PBS) can mix the carrier and modulated signal due to misalignment of reference axes as explained in the Fig.\ref{pol}. 
Same phenomena can be explained for PBC also. Following equations represents the effect of PBS angle ($\theta$) on the outputs (${PBS_x},{PBS_y}$):
\begin{equation*}
\begin{pmatrix}
{PBS_x} \\ {PBS_y}
\end{pmatrix}
= 
\begin{pmatrix}
\cos\theta & -\sin\theta  \\ \sin\theta & \cos\theta 
\end{pmatrix}
\begin{pmatrix}
{E_x} \\ {E_y}
\end{pmatrix},
\end{equation*}
\begin{eqnarray*}
	PBS_x=E_x \cos \theta - E_y \sin \theta,\\
	PBS_y=E_x \sin \theta + E_y \cos \theta,
\end{eqnarray*}
where ${E_x}, {E_y}$ are the inputs to the PBS. Phase shift between two orthogonal polarizations due to fiber channel can also cause the mixing. $\phi$ is the angle between the reference polarizations and the principle state of polarizations (PSPs). Overall effect due to the angles of PBS, PBC and $\phi$ can be represented as:
\begin{equation*}
A=
\begin{pmatrix}
\cos\theta & -\sin\theta  \\ \sin\theta & \cos\theta 
\end{pmatrix}
\begin{pmatrix}
e^{j\phi} & 0 \\ 0 & e^{-j\phi} 
\end{pmatrix}
\begin{pmatrix}
\cos\theta & \sin\theta  \\ -\sin\theta & \cos\theta 
\end{pmatrix},
\end{equation*}
Here assumption is that device angles of PBS and PBC are same.



\section{Separation of the carrier and the modulating signal at receiver:}
This section discusses the technique for separating the carrier and the modulated signal for the polarization diversity based SH-QPSK system. In this system as shown in Fig.\ref{block}, the modulated signal is launched in one polarization and the carrier signal is launched in other orthogonal polarization. Power of the launched carrier signal should be higher than power of the launched modulated signal. The power difference is kept around 15 dB at the transmitter (which is there due to modulator insertion loss). To separate the modulated signal and the carrier signal at the receiver, power measurement in one of the polarizations can help.
\begin{figure}[!h]
	\centering{
		\includegraphics[scale=0.80]{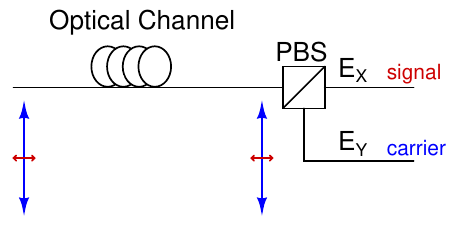}
		\includegraphics[scale=0.80]{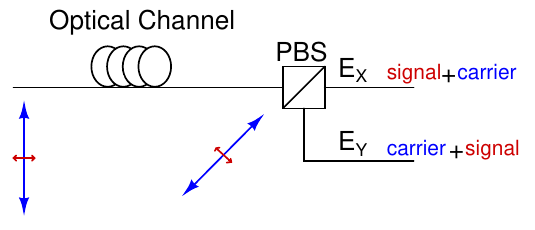}
		\caption{Effect of polarization control on the separation of signal and the carrier at the receiver.}
		\label{LS}		
	}
\end{figure}  
\par Under ideal conditions, the PBS has two outputs, the low power modulated signal and the high power carrier as shown in Fig.\,\ref{LS}. While, in practical scenario, these two signals may get mixed resulting in lower power difference between two output branches of the PBS. Thus, to extract the signal back in one polarization, a power minimization in that polarization at the receiver can be used. The polarization diversity based SH-QPSK system is modeled in Simulink and VPItransmissionMaker$^{TM}$ with three waveplate polarization controller (PC) module which has two control parameters for changing angles. Polarization impairments due to system and channel were considered non-ideal.  By varying the two controls of PC, optical power in one of the output of the PBS is measured and plotted. 
\begin{figure}[!h]
	\vspace{-0.4cm}
	\centering
	\includegraphics[scale=0.45]{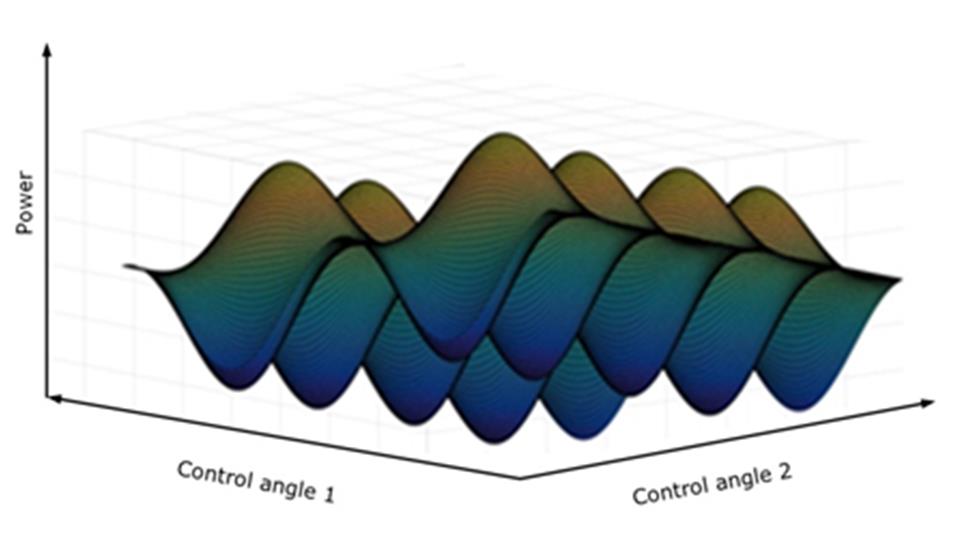}
	\vspace{-0.4cm}
	\caption{Optical Power profile from one of the output of PBS.}
	\label{pol3}
\end{figure}
\vspace{-0.2cm}
\par From the Fig.\,\ref{pol3}, it is shown that there are more than one minima with same strength. At each minima, equivalent matrix of the channel is converged to following matrix.
\begin{equation*}
\begin{pmatrix}
e^{j\phi1} & 0 \\ 0 & e^{-j\phi1} 
\end{pmatrix}
\end{equation*}
From this matrix, we can see that after minimizing the optical power in one of the polarization, rotational effect is removed and only phase shift in individual polarization is remaining. This phase shift can be removed using carrier and phase recovery module. Manual controlling of PC is difficult to get any one of the minima.  Feedback polarization control has been implemented in VPItransmissionMaker$^{TM}$ to get the desired state of polarization.  Optical power from any one of the polarization is converted in electrical domain using photodetector and based on the electrical signal it will adapt the control parameters of PC to find the minimum optical power.

\section{Adaptive polarization control algorithm }
Discrete time based gradient descent algorithm has been used to find the minima and validated through simulation. Here C1 and C2 are control parameters of the PC, P is optical power in one polarization and $\mu$ is stepsize.
\begin{figure}[!h]
	\vspace{-0.4cm}
	\centering
	\includegraphics[scale=0.65]{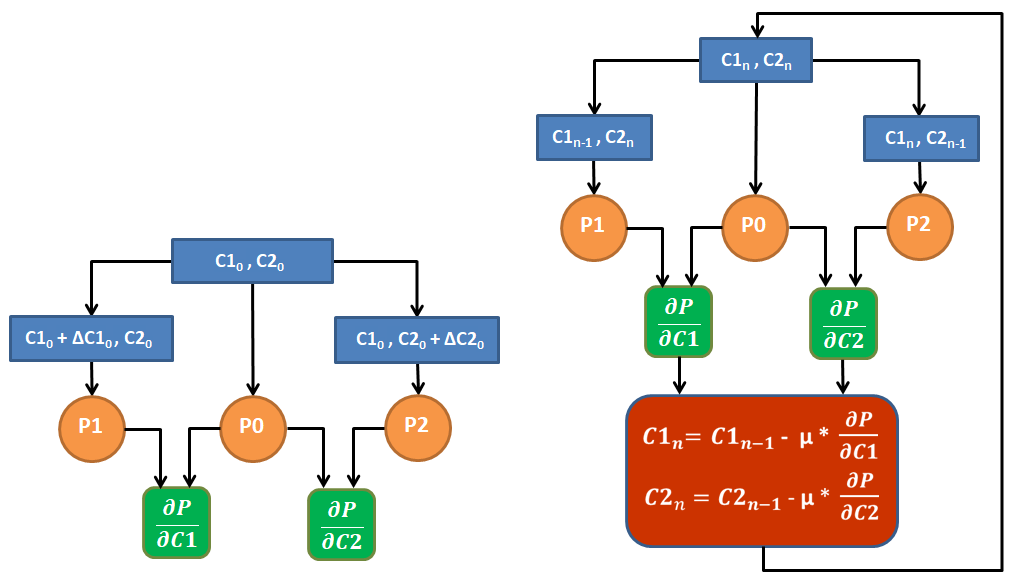}
	\vspace{-0.35cm}
	\caption{Flowcharts of algorithm for minimizing power}
	\label{pol3}
\end{figure}
\vspace{-0.4cm}
\section{Simulation results:}
\begin{figure}[!h]
	\vspace{-0.2cm}
	\centering
	\includegraphics[scale=0.25]{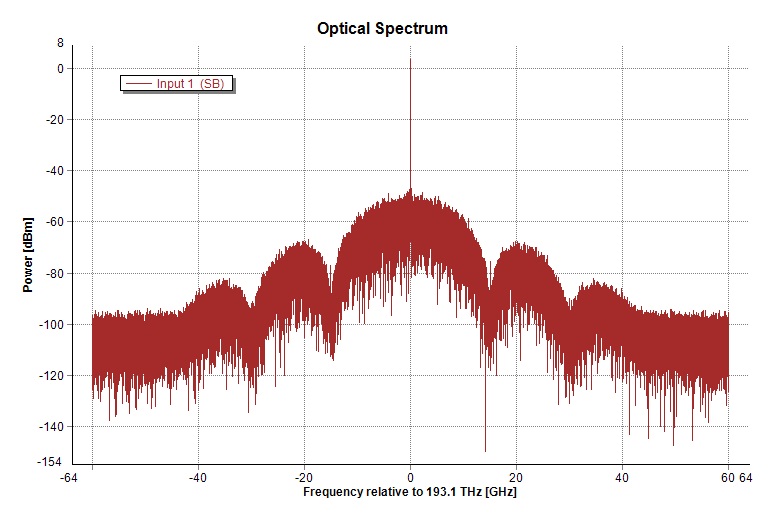}
	\caption{Date rate:30 Gbps, Distance: 20 km, Without minimizing power in one polarization: X pol.}
	\label{pol5}
\end{figure}

\begin{figure}[!h]
	\vspace{-0.2cm}
	\centering
	\includegraphics[scale=0.25]{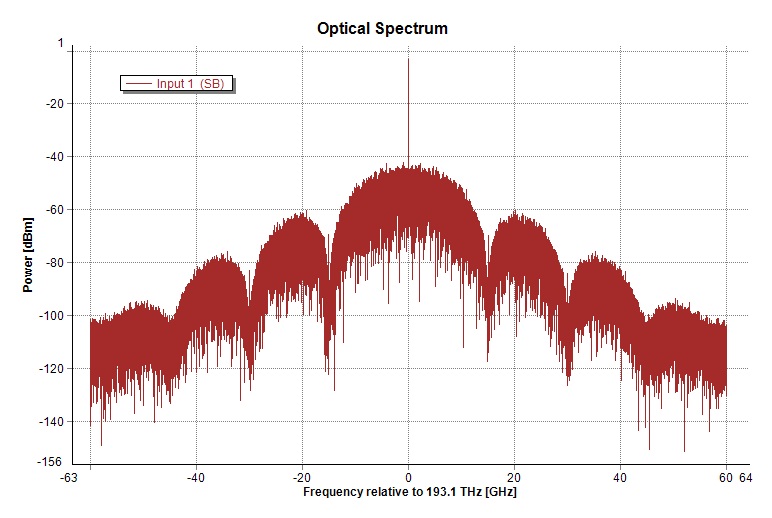}
	\caption{Date rate:30 Gbps, Distance: 20 km, Without minimizing power in one polarization: Y pol.}
	\label{pol6}
\end{figure}

\begin{figure}[!h]
	\vspace{-0.2cm}
	\centering
	\includegraphics[scale=0.25]{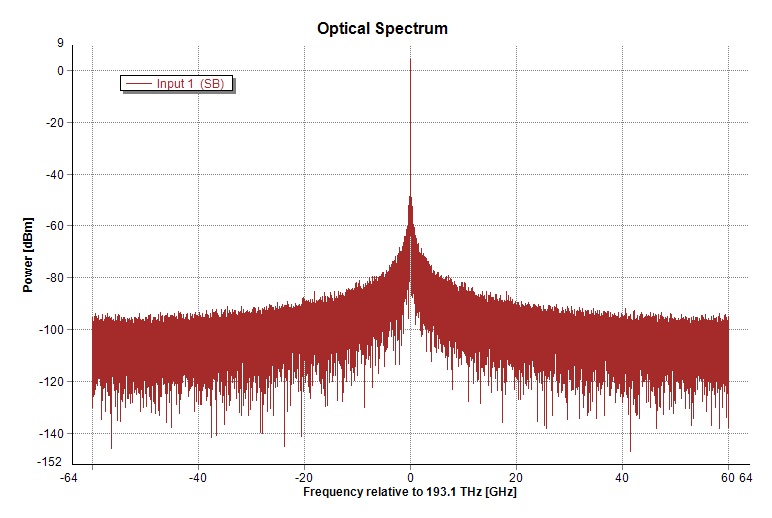}
	\caption{Date rate:30 Gbps, Distance: 20 km, With minimizing power in one polarization: X pol.}
	\label{pol7}
\end{figure}

\begin{figure}[!h]
	\vspace{-0.2cm}
	\centering
	\includegraphics[scale=0.25]{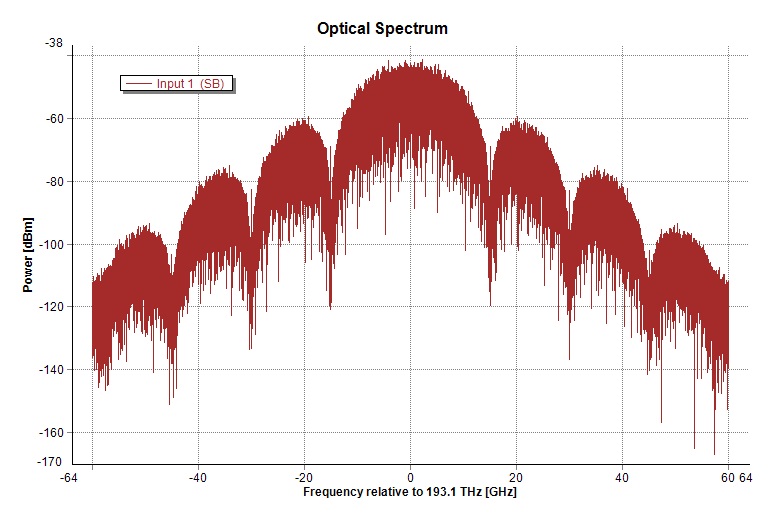}
	\caption{Date rate:30 Gbps, Distance: 20 km, With minimizing power in one polarization: Y pol.}
	\label{pol9}
\end{figure}

\begin{figure}[!h]
	\centering
	\includegraphics[scale=0.025]{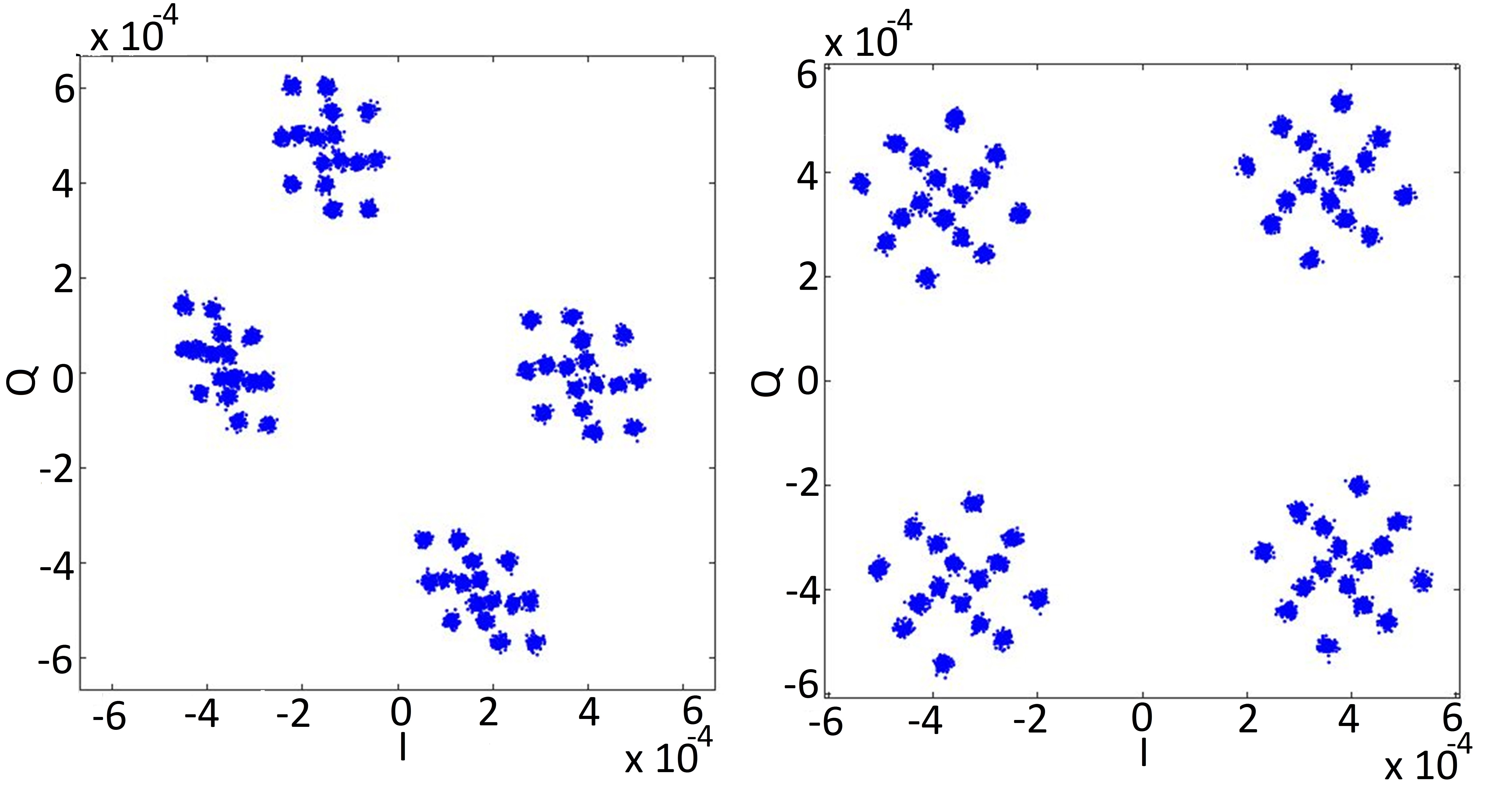}
	\caption{Constellation diagram of received signal without minimization of power (left) and with minimization of power (right).}
	\label{pol10}
\end{figure}

Simulation has been performed for 30\,Gbps SH-QPSK system with polarization multiplexed carrier with 20\,km distances in VPItransmissionMaker$^{TM}$ with all non-idealities on. Before minimizing power, power difference between two polarization was 6.04 dB (carrier and modulated signal are mixed) and after minimizing power in one polarization, power difference is 15.69 dB (carrier and modulated signal are separated). This is shown in  Fig.\,\ref{pol5}-\ref{pol10}. 
\par
\begin{figure}[!h]
	\centering
	\includegraphics[scale=0.2]{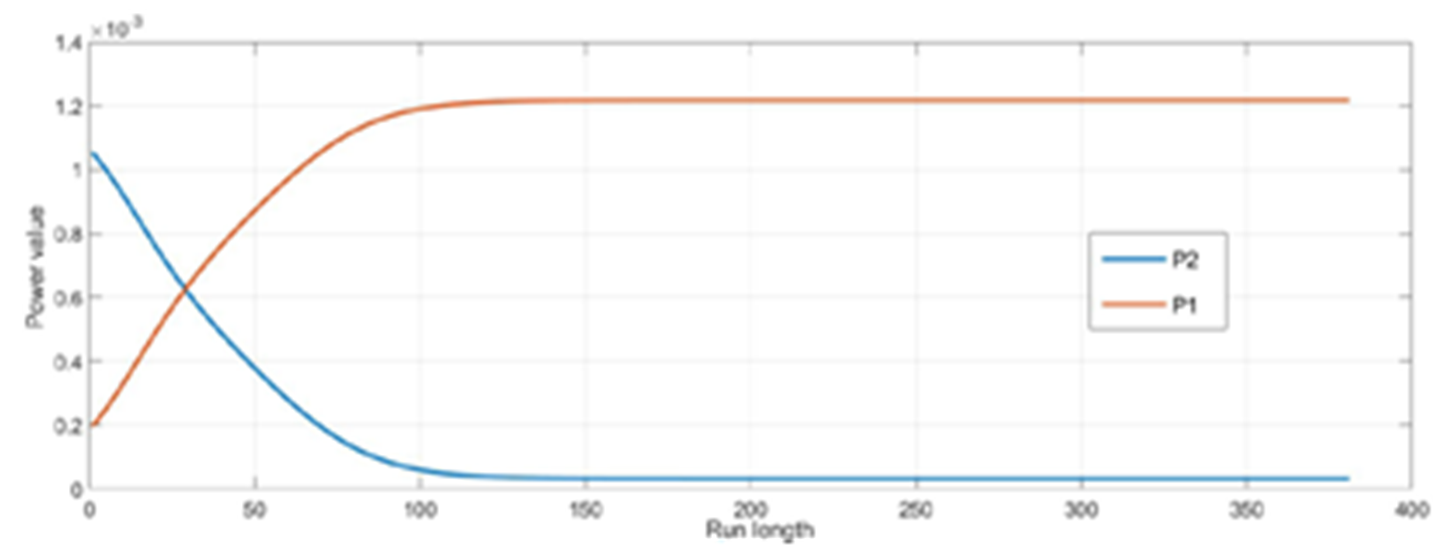}
	\caption{Power maximization in one polarization and minimization in other polarization by Gradient descent algorithm.}
	\label{pol8}
\end{figure}

 Power profile in X and Y polarization while applying gradient descent algorithm is also shown in in Fig.\,\ref{pol8}. Chromatic dispersion increases with data rate and fiber length. Chromatic dispersion does not cause the mixing of the carrier and the modulated data. It is shown that minimization of the power in one polarization is able to separate the carrier and the modulated signal even if there is a significant amount of chromatic dispersion. Due to chromatic dispersion, received data will be dispersed (Fig.\,\ref{pol10})

\section{Experimental results:}
\begin{figure*}[!h]
	\centering
	\includegraphics[scale=0.6]{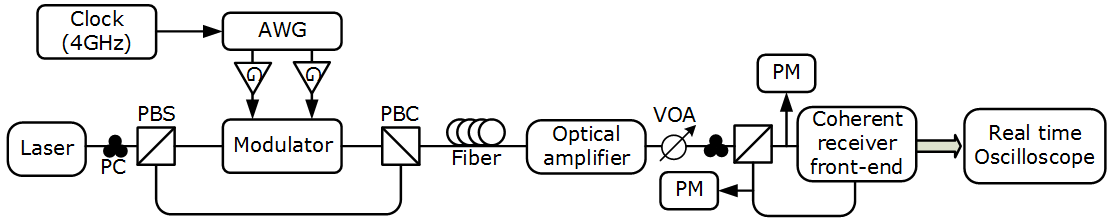}
	\caption{Experimental setup for SH-QPSK system with polarization multiplexed carrier. AWG: arbitrary waveform generator, PC: polarization controller, PBS/PBC: polarization beam combiner/splitter, VOA: variable optical attenuator, and PM: power meter. }
	\label{exp}
\end{figure*}
Block diagram of experimental setup of polarization diversity based SH-QPSK system is shown in  Fig.\,\ref{exp}. In this setup manually rotatable three paddle based PC is used to minimize the power in one of the polarization. SFL1550P external cavity laser (ECL) having power of 13.02 dbm is split into two polarization using PBS. Output power of the PBS branches are set using PC connected to laser source.  One of the branch of PBS is connected to the QPSK Modulator LN86S-FC which is driven by  two RF signals (2 Gbps/4 Gbps) generated by arbitrary wave form generator. Optical modulator driver- HMC788LP2E has been used to provide enough strength to RF signals. Another branch of the PBS is directly connected with output of the modulator using PBC.The received signal from the single mode fiber is split into orthogonal polarization (X and Y) using PBS. PBS branch having lower power is connected to the signal port of the receiver front end -CPRV1222A and branch having higher power is connected to the LO port of the receiver front end. For faithful demodulation of the receive data, the carrier and the modulated signal should not be mixed. But due to non-idealites of optical setup and fiber channel, these two polarization signals get mixed. In this case, there will be less power difference between two polarization signals after splitting at the receiver side. If these signals are directly connected to the receiver front end, it is not possible to get proper electrical data unless there is some signal processing done on it. 
\par Manually rotatable three paddles based polarization controller is connected to the fiber output to separate the carrier and the modulated data. To fulfill this requirement, two power meters are connected to the output branches of the PBS at the receiver side as shown in the Fig.\,\ref{exp}. Now by rotating the paddles of the polarization controller manually, maximum power difference between two polarizations is achieved. Branch having minimum power is connected to the signal port of the receiver front end and branch having maximum power is connected to the local oscillator port of the receiver front end. Minimizing power in one branch, data are captured. 
\vspace{-0.3cm}
\begin{figure}[h!]
	\centering{
		\begin{tabular}{cc}
			\includegraphics[width=36mm,height=30mm]{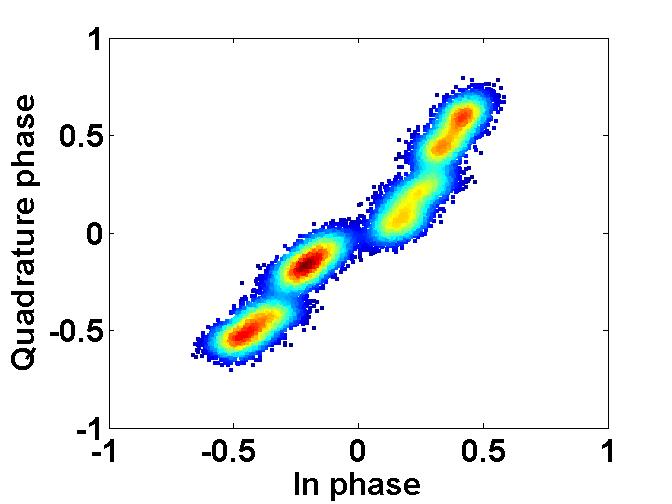}&
			\includegraphics[width=36mm,height=30mm]{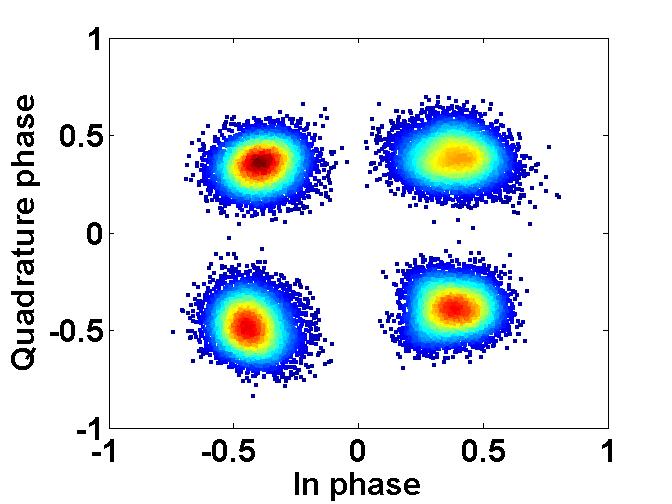}
			\vspace{-0.3cm} 
		\end{tabular}
		\caption{Data rate: 4 Gbps, Distance: 30 km, Constellation diagram: Received signal without minimiziing power (Power difference: 2.64 dB) (Left); Received Signal after power minimizing (Power difference: 13.2 dB, EVM after phase correction: 26.66 \%) (Right)}
		\vspace{-0.4cm}
		\label{dckm2}		
	} 
\end{figure}
\begin{figure}[!h]
	\centering
	\includegraphics[scale=0.3]{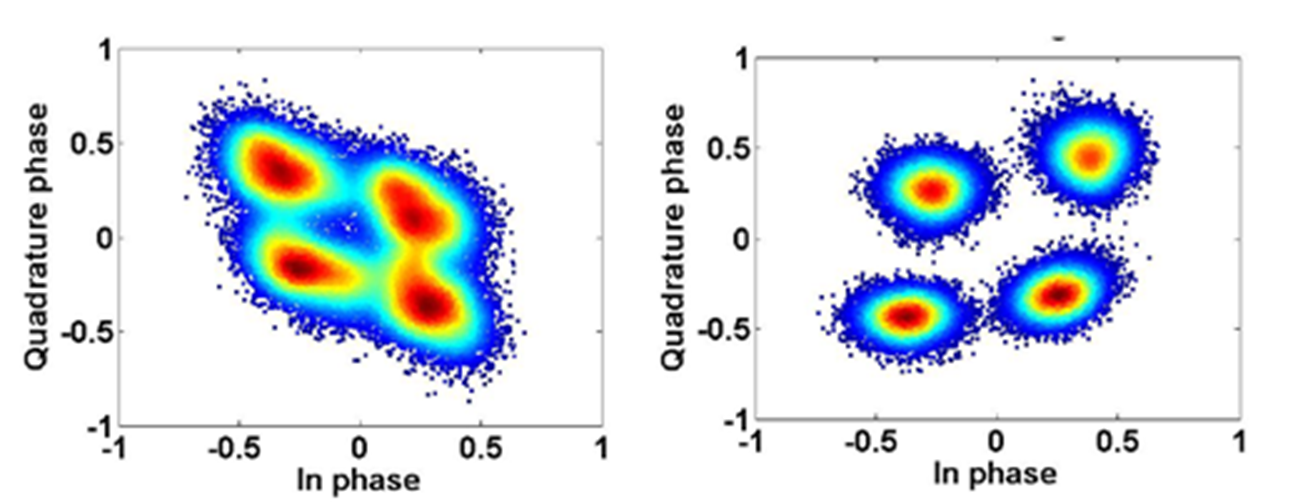}
	\caption{Data rate: 8 Gbps, Distance: 30 km, Constellation diagram: Received signal without minimiziing power (Power difference: 5.32 dB) (Left); Received Signal after power minimizing (Power difference: 13.03 dB, EVM after phase correction: 34.16 \%) (Right)}
	\label{pol11}
\end{figure}

Experiment has been performed for 30 km SSMF with data rate of 4 Gbps and 8 Gbps. Results are presented in Fig.\ref{dckm2} and Fig.\ref{pol11}. Improvement in the constellations is clearly observed in results with and without power minimization in one polarization.
\section{Conclusion} Proposed adaptive polarization control technique for polarization diversity based SH systems is practically feasible for implementation in real time systems. This proposed method is successfully validated with simulation and practical results. This work opens a choice to employ SH systems for low power high capacity data center interconnects.

\section*{Acknowledgment}
The authors would like to thank DST and DeitY for funding the project. We also like to thank Mr. Arvind Mishra and Mr. Madhan thollabandi from sterlite technologies for their continuous support.

\bibliographystyle{IEEEtran}
\bibliography{references}

\end{document}